# Toward the Theory of Electron and Positron States in Self-Compressed Dielectric Clusters

*Valentin V. Pogosov[1], Walter V. Pogosov[2], Denis P. Kotlyarov[1]*

[1] Department of Microelectronics, Zaporozhye National Technical University, Zhukovski Str. 63, Zaporozhye 69064, Zaporozhye, Ukraine

[2] Moscow Institute of Physics and Technology, Institutski per. 9, Dolgoprudny, Moscow Region, Russia

**Abstract**

Analytical expressions for the binding energy of electrons and positrons in dielectric clusters, analysed in this work, neglect the elastic effects. Therefore, we present the density-functional theory for neutral liquid clusters that experience the spontaneous deformation. Using the $1/R$-expansion, $R$ being the cluster radius, the exact analytical expressions for the size corrections to the chemical potential, surface tension, and atomic density are derived from the condition of mechanical equilibrium. The problem of calculation of these corrections is reduced to the calculation of the quantities for a liquid with a flat surface. The size compression and tension of density occur in the $1/R$ and $1/R^2$ orders respectively. The sizes of charged rigid and elastic critical clusters, for which the electron or positron binding energy is close to zero, are calculated for $Xe_N^-$, $Kr_N^-$, $Ar_N^-$, $Ar_N^+$, $Ne_N^+$, $He_N^+$. The calculations show significant contribution of self-compression to the binding energy of excess electron - in contrast to positron.

PACS: 36.40

## 1. Introduction

Excess charge particles and polarization interactions are of great importance in physical chemistry and biology. The interaction of electrons with atoms, which posses large polarizabilities, exhibits an attractive character. That is why their localization is possible in clusters [1-3]. Electronic clusters (or negative cluster's ions) were discovered experimentally in a dense xenon [4, 5]. In helium, which atomic polarizability is very small, localization of electrons happens in void bubble [6]. Recently, the electronic bubbles were observed even in the helium microdroplets [7]. The interaction of the positrons with atoms, owing to the absence of the exchange interaction, demonstrates always the attractive character. Positron clusters were discovered in all dense gases of rare atoms [8, 9]. The temperatures of clusterization and the "optimal" sizes of clusters were estimated in Ref. 2. Such clusters contain hundreds of atoms, and their density is close to that corresponding to liquid cluster. On the other hand, the mass-spectrometry measurements allowed to discover the existence of xenon clusters which contain near dozen of atoms and are charged by only one electron [10]. They have a noticeable lifetime and are called "critical" clusters. The size dependence of the electron affinity and critical size of xenon solid clusters were examined by continuum model [11], and by taking full account for the atomic structure [12]. In this work we propose an improvement of earlier theories.

The main purpose of this paper is to discuss a true asymptotic for binding energy of quantum particle localized in a large dielectric cluster. Subsequently, we point out the importance of elastic effects in the determination of the cluster's energetics. We develop a formal density-functional theory for finite classical system in order to account for the self-deformation of the clusters. For smallest clusters, the theory based on the continuum model retains the simplicity of the method developed for rigid clusters. Furthermore, critical sizes of single charged elastic clusters are calculated.

## 2. Large Rigid Cluster

The quantum particles localized in large clusters are almost free. Their energy spectra are determined by the character of scattering on cluster atoms and depend upon the atomic density. In Ref.13 the following expression for electron binding energy was discussed,

$$E_b = E_b^0 - \frac{\hbar^2 \pi^2}{2m_{eff} R^2}, \qquad (1)$$

where $E_b^0$ is the standard binding energy component that contains the Born correction,

$$E_b^0 = -V_0 - \frac{e^2}{2R} \frac{\varepsilon - 1}{\varepsilon}, \qquad (2)$$

where $V_0 < 0$ is the ground state energy of electron in a extended dielectric (Ar, Kr, Xe), $R = N^{1/3} \bar{r}$ is the cluster radius, $N$ is the number of its atoms, $\bar{r}$ is the average distance between the atoms of density $\bar{\rho} = \left(4\pi \bar{r}^3 / 3\right)^{-1}$. The second term in (1) is the kinetic energy of the electron localized inside cluster and $m_{eff}$ is the effective mass. The radius of critical clusters $R^*$ may be crudely estimated [14] directly from the condition $E_b^0(R^*) = 0$.

An alternative asymptotic expression for the binding energy of a charged particle has been derived in the effective medium approach and pseudopotential theory of scattering [15],

$$E_b = E_b^0 - \frac{\hbar^2 \pi^2}{2mR^2}(1 - C\xi), \qquad (3)$$

and

$$E_b^0 = T + \frac{3}{2} \frac{\alpha e^2}{\bar{r}^3 \sigma}\left(1 - \frac{\sigma}{R}\right) f, \qquad (4)$$

where $T = \mp \hbar^2 q_0^2 / 2m$. The sum of the first two terms in (4) gives $(-V_0)$ and the last term gives $-e^2(\varepsilon-1)/(2R\varepsilon)$. The dielectric constant $\varepsilon = 1 + 3\alpha/(\bar{r}^3 - \alpha)$ was taken in the Clausius-Mossotti approximation. The second term in (4) gives the shift of the energy due to the mean polarization of infinite liquid. The minus and plus sign appearing in $T$ correspond to $L > 0$ and $L < 0$, respectively, where $L \equiv L(\bar{r})$ is the scattering length of a quantum particle in dielectric. $\alpha$ is the atomic polarizability, $\sigma$ is the parameter of the Lennard-Jones potential, $f = (1 + 2\alpha \bar{r}^3)^{-1}$ is the Lorentz local-field correction, $C \approx 2.86$, and $\xi = L/\bar{r}$ is the small parameter. A simple form of step function was used for the pair-correlation function for atoms,

$$g(r) = \theta(\sigma - r), \tag{5}$$

where $\sigma$ corresponds to the mean closest interatomic distance in the cluster[1]. The solution of the Schrodinger equation in the Wigner-Seitz cell for the two principally different regimes of scattering [17], gives the following equation for $q_0$

$$\tan[q_0 \bar{r} + \delta_0(q_0)] = q_0 \bar{r}, \quad \delta_0 = -L q_0 + O(q_0^3), \text{ for } L > 0,$$
$$\tanh[q_0 \bar{r} + \operatorname{Im} \delta_0(iq_0)] = q_0 \bar{r}, \quad \delta_0 = -i L q_0 + O(-i q_0^3), \text{ for } L < 0. \tag{6}$$

Here $\delta_0(x)$ is the phase shift of the charged particle's $s$-wave scattered in cellular infinite medium[2].

In principle both expressions (1) and (3) follow from the Bardeen theory [19] for extended system. They give, however, different size-dependence of binding energy. In this section, our consideration are restricted to a special case of large clusters when both the electron mean free path in extended liquid (which is of the order of hundreds of bohrs) and the electron wave length in the cluster are close to cluster radius. In this case the binding energy should be calculated from Eq.(3). Calculations using Eq.(1), assuming the input of effective mass, are not correct because the effective mass can be correctly calculated and entered to (1) only if the mean free path is much smaller than the cluster radius.

We describe the fluid number density $\rho$ as of undisturbed fluid of uniform density up to spherical boundary, i.e. as for a *rigid* cluster with zeroth compressibility, and we put $\rho(r) = \bar{\rho}\theta(r - R)$. The values of $V_0$ and $m_{\text{eff}}$ for electrons and positrons in considered media were measured in a wide range of densities [20-26]. We calculated these values (Table 1), taking into account simple correlation function given by (5). The input experimental values of $L(\bar{r})$ for excess electrons are taken from Refs. 4, 25, 26. The input calculated values of scattering lengths for positrons are used from Ref. 17.

---

[1] In Refs.16, 17 the radial distribution function $g(r)$ was used, which reflected the real structure of simple liquids in coefficients $I_0$, $I_2$, $I_4$ appearing in the expressions for the phase shifts of scattering waves, $V_0$ and $m_{\text{eff}}$. The present version of $g(r)$ corresponds to $I_0 = I_2 = I_4 = 1$.

[2] It should be noted that $\hbar^2 q_0^2 / 2m$ appearing in (4) is not kinetic energy of particle in the cell, as it seems to be. This term describes only scattering inside a cell. The wave number $q_0$ is obtained from (6) using the boundary conditions by means of the scattering length which allow to account entirely for the repulsion and partially for the attraction, i.e., sattering at the polarization potential profile inside cell [18].

In Fig.1 the binding energies $E_b(N)$ calculated from (1) and (3) are shown for $Xe_N^- \equiv Xe_N + e^-$, $Kr_N^- \equiv Kr_N + e^-$ and $Ar_N^+ \equiv Ar_N + e^+$ clusters of densities corresponding to a liquid state in the triple point. As is seen from Fig.1, the two curves for $E_b(N)$ differ considerably. The difference in the curves for $Xe_N^-$ and $Ar_N^+$ originates from the effective masses $m_{eff}$ and from the sign of scattering length $L$. Equation (3) predicts smaller size of "critical" cluster $Xe_N^-$ and $Kr_N^-$ (which correspond to the condition $E_b(N^*) = 0$). These results suggest that Eq. (3) is superior over (1) because it predicts smaller sizes of critical electronic clusters. In Section 4 we show that these sizes are determined by the availability of surface states. Latter effect was ignored in Refs. 14, 15.

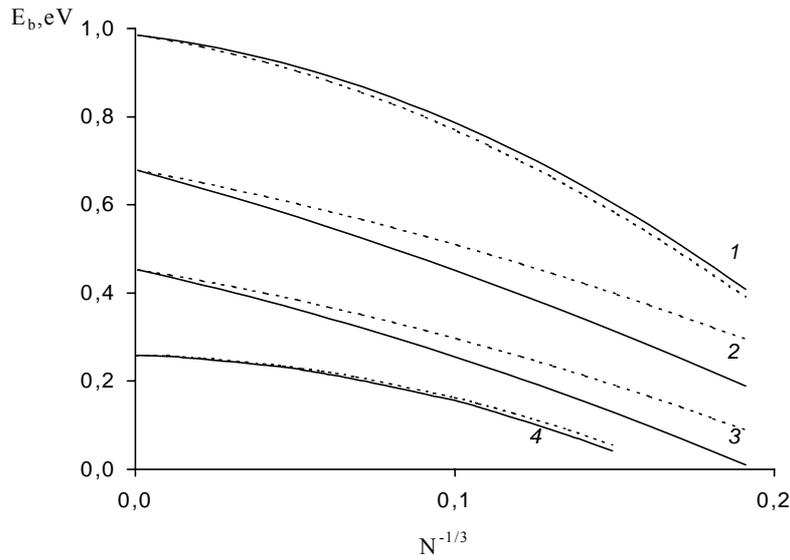

**Fig. 1**. The binding energy $E_b(N)$ calculated from Eqs.(1) and (3) (dashed and solid line, respectively) for: 1 - $Ar_N^+$; 2 - $Xe_N^-$; 3 - $Kr_N^-$.

Finally, putting aside the problem of availability of surface states, we should add that formula (3) is formally correct but exclusively for electronic clusters, $L > 0$, with N>100. This is confirmed by numerical solution of eigenvalue problem for the potential well of radius R and of depth $E_b^0$. The point is that the sum of exact kinetic energy and the last term that contains weight coefficient $C \equiv C(R)$ (see [15]) shows the size dependence similar to that for large clusters in Eq. (3).

However, in general, the cluster may be compressed under the action of surface tension and tensed by a localized quantum particle. We can neglect the pressure $P_q$ of localized charge $q = \pm e$ in two cases: for a large and "bulk" cluster ($P_q$ has an order $R^{-4}$, that is much less then the Laplace pressure) and for a critical cluster ($E_b \to 0$, $P_q \to 0$). In these cases one can take into consideration, in analytical form, the effect of self-compression of cluster under the action of surface forces upon the energetics of a bound quantum particle. For intermediate sizes of clusters, the self-consistent solution of the problem of a particle localized in the liquid cluster is required.

The analytical sum-rule approach, developed for neutral metallic clusters [27, 28], describes the influence of self-compression upon ionization potential only in terms proportional to first order in $1/R$. As will be shown, for dielectric cluster this approach is more progressive, and the desired corrections proportional to $1/R^2$ are obtained. In the following section we briefly present the density-functional theory of self-deformed cluster.

## 3. Density-Functional Theory

Consider a classical, dense vapour at temperature $T$, and of chemical potential $\mu$, in a box of volume $V$. The free energy of a system of cluster-vapour, $F \equiv F[\rho(r,R)]$ is a functional of the inhomogeneous atomic concentration $\rho(r,R)$, $R$ is the cluster radius. In the framework of the square-gradient theory, the free energy can be written in the form

$$F = \int d^3 r \left( f + g (\nabla \rho)^2 \right), \qquad (7)$$

where $f \equiv f[\rho(r,R)]$ is the energy density of the quasi-homogeneous part of the functional, $g \equiv g[\rho(r,R)]$ gives the first inhomogeneity term represented by the first gradient term.

The grand free energy is found by minimizing the functional

$$\Omega_V [\rho] = F[\rho] - \mu \int d^3 r \rho(r) + \int d^3 r \frac{\hbar^2}{2m} |\nabla \psi|^2 + \int d^3 r \int d^3 r' |\psi(r)|^2 v(r - r') \rho(r') \qquad (8)$$

with respect to variation of $\rho(r)$ and $\psi(r)$ under the conditions

$$\int d^3 r |\psi(r)|^2 = 1, \quad \int d^3 r \rho(r) = N_0.$$

Here $v(r)$ is the electron/positron – atom potential, and $N_0$ denotes the total number of atoms in a box. By varying $\Omega_V [\rho]$ with respect to $\psi(r)$ while using a Lagrange multiplier, one finds the following Schrodinger equation for $\psi(r)$:

$$-\frac{\hbar^2}{2m} \Delta \psi(r) + V(r) \psi(r) = E \psi(r), \qquad (9)$$

where

$$V(r) = \int d^3r' v(r-r')\rho(r') \qquad (10)$$

is the mean potential field, produced by atoms. For a given $\rho(r)$ we want the lowest-energy solution to Eq.(9). Let denote the energy in this state by $E[\rho]$. For the equilibrium profile $\rho(r,R)$, the functional $\Omega_V[\rho(r,R)] = E + F - \mu N_0$ has a minimum and equals the Gibbs grand potential $\Omega = -PV$, where P is the pressure in a box.

In this paper we use $V(r)$ in the form of the sum of short range (see Eq.(4)) and long range (polarization) components:

$$V(r) = T\delta(r) + \int d^3r' V_p(r-r')\rho(r'), \qquad (11)$$

where $\delta(r)$ is the Dirac $\delta$ function. For dense cluster in the delute vapour the last term in (11) has a standard form of the interaction energy of a point charge with a dielectric sphere [29,16]. We consider the case of weak perturbation of the atomic distribution $\rho(r)$ by excess quantum particle (see the above discussion in Section 2). However, the effect of the correction may be estimated after the fact and such an estimate is made in the end of the Section.

### 3.1. Neutral elastic cluster

Using (7), the Euler-Lagrange equation can be written in the form

$$\mu(r,R) = \frac{\delta F[\rho]}{\delta \rho(r,R)} = \frac{\partial f}{\partial \rho} - \frac{\partial g}{\partial \rho}(\nabla\rho)^2 - 2g\left(\Delta\rho + \frac{2}{r}\nabla\rho\right). \qquad (12)$$

For the equilibrium concentration profile $\rho(r,R)$ we have $\mu(r,R) \equiv \mu(R)$. By definition, the surface free energy per unit area, $\gamma$, and surface tension (stress, for a solid) $\tau$ [30] are given by

$$\gamma = \frac{1}{A}\left[F[\rho(r,R)] - F[\rho_0^+]\theta(r-R) - F[\rho_0^-]\theta(R-r)\right], \qquad (13)$$

$$\tau = \gamma + A\frac{d\gamma}{dA}, \qquad (14)$$

where $A = 4\pi R^2$ is the area of "equimolecular surface" of cluster, which is defined by the condition

$$4\pi\int_0^\infty dr r^2 \left(\rho(r,R) - \rho_0^+\theta(r-R) - \rho_0^-\theta(R-r)\right) = 0. \qquad (15)$$

Here $\rho_0^+$ is the atomic concentration in the uniform condensed matter, $\rho_0^-$ the density of uniform vapour beyond the surface, and $\theta(-x)$ the Heaviside step function. In following we employ the expansion of $Y \equiv \rho, \mu, \gamma, \tau$ quantities in powers of the inverse radius $1/R$,

$$Y = \sum_{k=0}^\infty \frac{Y_k}{R^k}. \qquad (16)$$

The zeroth-order terms in (16) are relevant to the system with planar boundary. Inserting this expansion into (12) and (14), and using the series

$$\frac{1}{r} = \frac{1}{R}\sum_{k=0}^{\infty}(-1)^k\left(\frac{r-R}{R}\right)^k,$$

one can compile the terms having equal powers of $1/R$, getting a set of equations for $\rho_k$ and $\mu_k$. The equations for $k = 0,1,2$ have the form

$$\mu_0 = \frac{\partial f_0}{\partial \rho_0} - \frac{\partial g_0}{\partial \rho_0}(\nabla \rho_0)^2 - 2g_0 \Delta \rho_0, \quad (17)$$

$$\mu_1 = \frac{\partial^2 f_0}{\partial \rho_0^2}\rho_1 - 2\frac{\partial g_0}{\partial \rho_0}(\nabla \rho_0 \nabla \rho_1 + \rho_1 \Delta \rho_0) - \frac{\partial^2 g_0}{\partial \rho_0^2}\rho_1 (\nabla \rho_0)^2 - 2g_0(\Delta \rho_1 + 2\nabla \rho_0), \quad (18)$$

$$\mu_2 = \frac{1}{2}\frac{\partial^3 f_0}{\partial \rho_0^3}\rho_1^2 + \frac{\partial^2 f_0}{\partial \rho_0^2}\rho_2 + \text{gradient terms}, \quad (19)$$

$$\int_{-\infty}^{\infty}dx\left(\rho_0 - \rho_0^+ \theta(-x) - \rho_0^- \theta(x)\right) = 0, \quad (20)$$

$$\int_{-\infty}^{\infty}dx\left(\rho_1 - 2x\rho_0^+ \theta(-x) - 2x\rho_0^- \theta(x)\right) = 0, \quad (21)$$

where we have changed the variable $x = r - R$, and we have made use of the limit $R \to \infty$, $\rho^+ \equiv \rho(x = -\infty)$, $\rho^- \equiv \rho(x = +\infty)$. For brevity, we use the notation $\nabla = d/dx$, and $\Delta = d^2/dx^2$. The liquid under consideration occupies the half-space $x < 0$, and vapour is for $x > 0$. It is convenient to introduce the useful definition of the "average over a planar surface"

$$\langle \mu(x) \rangle \left(\rho_0^+ - \rho_0^-\right) = -\int_{-\infty}^{\infty}dx\mu(x)\nabla \rho_0, \quad (22)$$

and "first average over spherical surface"

$$\langle\langle \mu(x) \rangle\rangle \left(\rho_1^+ - \rho_1^-\right) = -\int_{-\infty}^{\infty}dx\mu(x)\nabla \rho_1. \quad (23)$$

To transform in Eq.(13) we have to carry out the following procedure. Multiply Eq.(17) by $\nabla \rho_0(x)$ and then express the result in the form

$$\nabla\left(f_0 - g_0(\nabla \rho_0)^2 - \mu_0 \rho_0\right) = 0 \quad (24)$$

which represents a microscopic analogue of the condition of mechanical equilibrium for cluster-vapour system. Next, integrate Eq.(24) in the limits $(-\infty, x)$ to yield

$$f_0(x) = f_0^+ + g_0(\nabla \rho_0)^2 + \mu_0(x)\rho_0(x) - \mu_0^+ \rho_0^+, \quad (25)$$

where $f_0^+ \equiv f(\rho_0^+)$. It makes possible to separate $\gamma_0$ and $\gamma_1$ in the expression (13) for $\gamma(R)$. Using Eqs. (20)-(22), after cumbersome transformations, one gets the analogue of results obtained earlier, and in another form, in the framework of the Van der Waals theory (see [31-33], where $g_0$ =constant was used)

$$\gamma_0 = 2\int_{-\infty}^{\infty} dx g_0 (\nabla\rho_0)^2, \qquad (26)$$

$$\gamma_1 = 4\int_{-\infty}^{\infty} dx x g_0 (\nabla\rho_0)^2. \qquad (27)$$

A similar expression was derived earlier [34] in the two-component plasma model and stabilized jellium for self-compressed metal cluster.

We conclude this section by deriving necessary exact sum-rules. Using Eqs.(17), (18), (22), (23), and (26), (27) one can obtain the following expressions,

$$\mu_1^\pm = \rho_1^\pm \frac{\partial^2 f_0^\pm}{\partial \rho_0^{\pm 2}}, \qquad (28)$$

$$\langle\mu_1\rangle(\rho_0^+ - \rho_0^-) = 2\gamma_0, \qquad (29)$$

and

$$\mu_2^\pm = \frac{1}{2}\rho_1^{\pm 2} \frac{\partial^3 f_0^\pm}{\partial\rho_0^{\pm 3}} + \rho_2^\pm \frac{\partial^2 f_0^\pm}{\partial\rho_0^{\pm 2}}, \qquad (30)$$

$$\langle\mu_2\rangle(\rho_0^+ - \rho_0^-) + \frac{1}{2}\langle\langle\mu_1\rangle\rangle\left(\frac{\rho_0^{+2}\mu_1^+}{B_0^+} - \frac{\rho_0^{-2}\mu_1^-}{B_0^-}\right) = 2\gamma_1, \qquad (31)$$

where $\mu^+ \equiv \mu(x=-\infty)$, $\mu^- \equiv \mu(x=+\infty)$, and $B_0^\pm = \rho_0^{\pm 2}\partial^2 f_0^\pm/\partial\rho_0^{\pm 2} \equiv \rho_0^{\pm 2} f_0^{\pm''}$ is the bulk modulus (or inverse compressibility) of liquid and vapour, respectively. In particular, Eq.(29) defines the size correction to the "atomic work function" or cohesive energy $\varepsilon_{coh}(R) = \varepsilon_{coh0} + \varepsilon_{coh1}/R$, where $\varepsilon_{coh1} = -2\gamma_0/(\rho_0^+ - \rho_0^-)$.

The equilibrium conditions, $\mu_{1,2}^+ = \langle\mu_{1,2}\rangle = \mu_{1,2}^- = \langle\langle\mu_{1,2}\rangle\rangle$, lead to cancellation of the second term in (31) and after trivial algebra we derive the desired equalities

$$\rho_1^+ = 2\gamma_0 \frac{\rho_0^{+2}}{B_0^+(\rho_0^+ - \rho_0^-)}, \qquad (32)$$

$$\rho_2^+ = \rho_1^+(\delta - \chi), \qquad (33)$$

which will be used in further calculations. The "size" coefficient $\delta = \gamma_1/\gamma_0$ is defined by the dependence $\gamma(R) = \gamma_0(1 + \delta/R)$, and $\chi = \rho_1^+ f_0^{+'''}/2f_0^{+''}$. The quantities $\rho_1$, $\rho_2$ appearing in (32) and (33) can be calculated by solving the problem for a flat surface. It should be noted that for liquid rare gases the value $\gamma_0 / B_0^+$ is close to one half of the Bohr radius $a_0 = \hbar^2 / me^2$, thus giving some "fundamental" length by analogy with the liquid metals [34].

Expression (32) means that atomic concentration in the bulk of the cluster increases by $\rho_1^+ / R$ compared to the $\rho_0^+$ case where $R \to \infty$. Thus, *self-compression* is a result of surface curvature that creates extra pressure, $2\gamma_0 / R$, in comparison to the planar case. It will be demonstrated below, that the second next correction, $\rho_2^+ / R^2$, has a negative sign. This points to the size *self-tension* that appears in the term of order $1/R^2$.

The sign of coefficient $\delta$ in (32) may be derived intuitively in the following way. The response of cluster to decreasing of it size corresponds to the well-known Le Chatelier principle. Taking into account the size dependence of surface energy, the extra pressure inside the cluster is $2\gamma(R)/R$, where $\gamma(R) < \gamma_0$. Consequently, the decreasing of $\gamma(R)$ in comparison to $\gamma_0$, counteracts the increase in capillary pressure, caused by the decreasing of cluster size. In order to make a connection to Ref. 10, we will restrict our consideration to the cluster-vacuum system. It means that we need to set $\rho_0^- = 0$ in (32) and to make a change $\gamma(R) \to \tau(R)$ [35]. Then, by definition of surface tension (14), we have

$$\tau(R) = \tau_0 \left(1 + \frac{\delta}{2R}\right). \tag{34}$$

Here, for simplicity we assume that $\tau_0 = \gamma_0$ (see discussion in Refs. 36-38). Following Eq. (34), the correction $\rho_2^+ / R^2$, defined by (33), decreases by a factor of 2.

Let's discuss the influence of localized quantum particle upon the atomic density in a cluster. In general, the corresponding component of pressure is defined by two last terms of Eq.(8). In the consider system the intrinsic pressure has a form

$$P = \frac{2\tau(R)}{R} + P_q[\psi(r), \rho(r)]. \tag{35}$$

The pressure $P_q$ stipulated by excess particle is defined by derivative $dE/dV_{cl}$ over the volume of cluster $V_{cl}$, $E = -E_b$. For large cluster, $\int_0^R dr 4\pi r^2 |\psi(r)|^2 \to 1$, $E_b$ corresponds to Eqs.(2),(3), and this component of pressure can be written as follow

$$P_q \to \left\{ -\frac{e^2}{8\pi R^4} \frac{\varepsilon - 1}{\varepsilon} - \frac{\hbar^2 \pi}{4mR^5} (1 - C\xi) \right\}. \tag{36}$$

Thus we obtain an analogue of the Tompson equation [39]. The "surplus" pressure $P_q$ of quantum particle introduces the additional correction to atomic density $\Delta \rho_q = \rho_0 P_q / B_0^+$ (see Section 1). Simple calculations demonstrate the weak effect of tension of Xe clusters, induced by a particle in the range for $N > 100$. With decreasing $N$ it effect becomes some noticeable. However, for smallest, i.e., near critical clusters, this physical picture become simpler, because occupation probability for electron (or positron) is close to zero, and the pressure term $P_q$ disappear.

## 3.2. Small clusters.

Consider the ground state of particle localized in a small cluster. Usinq (8), let us write the wave equation for the radial wave function

$$\frac{d^2 u(r)}{dr^2} - \frac{2m}{\hbar^2}[E_b + V(r)]u(r) = 0, \qquad (37)$$

where $u(r) = r\psi(r)$, $\psi(r)$ is the particle wave function and the potential $V(r) \equiv V(R, r)$. The ground–state wave function is symmetrical about the centre of cluster, so that the boundaries conditions $u(0) = 0$, and $u(\infty) = 0$, have to be satisfied.

With cluster size decreasing the near-surface region occupies the considerable part of its volume and electron mainly can be found to be outside formal cluster boundary at the polarization tail of the potential $V(r > R)$. It is stipulated by the electrostatic component of $V(r)$, which can be calculated exactly as the interaction energy of a point charge $e^\pm$ with the dielectric sphere of radius $R$. The behaviour of the electrostatic component of $V(r)$ at boundary has unphysical singularity [29]. Therefore, the singularity at $r = R$ is removed by a usual cut-off procedure and replaced by a constant potential. The discontinuities of $V(r)$ are an artefact of this model and have only a small influence on $\psi(r)$ and the binding energy [11]. On the other hand, the short-range component of $V(r)$ can be calculated only when $r \leq R$ [15] (see Eq.(11)). Thus, we assume that the one-particle "pseudopotential" in Eq. (37) has a form similar to the Heine-Abarenkov electron-ion pseudopotential for a metal, i.e. it can write as follows

$$V(r) = \begin{cases} -E_b^0, r < R \\ V_p(R + \bar{r}/2), R < r < R + \bar{r}/2 \\ V_p(r), r > R + \bar{r}/2 \end{cases}, \qquad (38)$$

where for the polarization tail $V_p(r)$, the cut-off at $r = R + \bar{r}/2$ is used, and

$$V_p(r) = -\frac{e^2}{2}\frac{\varepsilon-1}{\varepsilon+1}\frac{R}{r^2}\left[\frac{R^2}{r^2 - R^2} - \frac{1}{\varepsilon+1}\left[\ln\left(\frac{r^2}{r^2 - R^2}\right) - \sum_{k=1}^{\infty}\frac{1}{k(k\varepsilon + k + 1)}\left(\frac{R}{r}\right)^{2k}\right]\right], r > R; \qquad (39)$$

The pseudopotential (38) has the right asymptotics: $V(r) \to V_0$ for $\sigma/R \to 0$, and $V(r) \to -N\alpha e^2/2r^4$ for $r/R \to \infty$. The binding energy $E_b$ results from a competition of kinetic and polarization energies, and for a critical cluster is close to zero. Thus, solving Eq. (37), we find $V(R^*, r)$, and consequently $N^* = (R^*/\bar{r})^3$.

The potential in the centre of large cluster can be assumed as the nearest to the bottom $V_0$ of the conduction-band in the infinite liquids. For solid state, $V_0$ is close to zero (especially for Ar) [20], and by taking into account the Born size correction and self-compression it becomes even positive (more incapable to retain an electron). On the other hand, polarization tail $V_p(r)$, in the region $r > R$, depends rather weakly upon cluster state (liquid or solid). Therefore it is clear, that when the first bound state appears, the electron will probably be localised outside the cluster, in a near-surface state.

For positron, in contrast to electron, it is more probable that it will be situated inside the cluster. In large cluster $Ar_N$ the value $V_0$ is about -1eV that is, in the centre of the cluster, positron feels a deep potential well. Positron localizes on much smaller clusters of Ar than electron. This is conditioned by the comparative prevalence of attraction over repulsion in the positron-atom interaction.

## 4. Calculations
### 4.1. Large clusters

First, let us define $\delta = c_1 + c_2$ for the calculation of $\rho_2^+ = \frac{1}{2}\rho_1^+(\delta - \chi)$ (see Eq. (33) and comment below Eq. (34)). From the semi-empirical rule [38], derived from the vacancy formation energy and the cohesive energy results $c_1 = +0.5\bar{r}_0$. The re-definition of "equimolecular surface" for icosahedral cluster [36] gives $c_2 = -1.32\bar{r}_0$ and thus $\delta = -0.82\bar{r}_0$. The calculation of third derivative of free energy with respect to density is difficult problem. On the other hand, the third derivative in (33) can be expressed by the first derivative of the $(B_0^+/\rho_0^{+2})$ with respect to $\rho_0^+$. Let us use the well-known sum-rule for compressibility

$$S_{k=0} = \frac{\rho_0^+ k_B T}{B_0^+}, \qquad (40)$$

where $S_{k=0}$ is the structure factor of a liquid for zeroth wave vector and for constant temperature, T. For bulk properties of liquids the hard-sphere model gives good results so, we employed Percus-Yevick fluid structure factor, $S_{HS} = (1-\eta)^4/(1+2\eta)^2$. Here, $\eta = \pi d^3 \rho_0^+/6$ is the packing fraction and d is the hard-sphere diameter. Then, we have $\chi = (\gamma_0/B_0^+)(\rho_0^+/y)(\partial y/\partial \rho_0^+)$, where $y = k_B T/S_{HS}\rho_0^+ = B_0^+/\rho_0^{+2}$. Using the experimental magnitudes of $S_{k=0}$ in triple point [40] we determine d and then $\chi$. This allows rewriting expression (33) with a reasonable accuracy in the following form

$$\rho_2^+ = -\varsigma \bar{r}_0 \rho_1^+, \qquad (41)$$

where $\varsigma$ is the constant (see Table 1). Comparing the values of $\rho_1^+$ and $\rho_2^+$ one can see that size tension is noticeable effect on the atomic density corresponding to smallest clusters.

**Table 1**. *The calculated input values of $V_0$ and $m_{eff}$, and used for estimation of the binding energy $E_b(N)$. The data are taken from [4,25,26,17,40, 41], $a_0$ is the Bohr radius.*

|  | T [K] | $\bar{r}_0$ [$a_0$] | $L(\bar{r}_0)$ [$a_0$] | $V_0$ [eV] | $dV_0/d\rho$ [eV×$a_0^3$] | $m_{eff}/m$ | $\gamma_0/B_0^+$ [$a_0$] | $\varsigma$ |
|---|---|---|---|---|---|---|---|---|
| $Xe_N^-$ | 161.4 | 4.855 | 0.70 | -0.680 | +1140 | 0.664 | 0.63 | 1.10 |
| $Kr_N^-$ | 115.7 | 4.544 | 0.60 | -0.454 | +676 | 0.678 | 0.57 | 1.02 |
| $Ar_N^-$ |  |  | 1.10 | -0.201 | +1122 | 0.711 |  |  |
|  | 83.8 | 4.225 |  |  |  |  | 0.49 | 0.97 |
| $Ar_N^+$ |  |  | -0.63 | -0.986 | -303 | 1.203 |  |  |
| $Ne_N^+$ | 24.8 | 3.531 | -0.027 | -0.446 | -17.8 | 1.099 | 0.46 | 0.93 |
| $He_N^+$ | 4.2 | 4.404 | -0.29 | -0.259 | -45.0 | 1.05 | 0.44 | 0.95 |

It should be noted that the compression of cluster leads to the rise/drop of potential bottom $V_0$ for electron/positron and to growth of its kinetic energy owing to the decrease of radius. The position of the bottom of the band shows strong dependence on the density of atoms (see [17] and Table 1). In following, for simplicity calculation of the $E_b^0(N)$ component in (3), for self-deformed clusters we employed linear approximation

$$V_0(\rho) = V_0(\rho_0) + \frac{dV_0}{d\rho}(\rho - \rho_0),$$

and

$$\rho - \rho_0 = \rho_1^+/R_0 + \rho_2^+/R_0^2,$$

where, $R_0 = N^{1/3}\bar{r}_0$. As illustration, Fig. 2 compares the electron binding energy, calculated from Eq. (3), for elastic and for rigid clusters. The difference is much greater than the energy $k_BT$ of thermal excitation. One can see that the shrinkage of a Xe or Kr cluster leads to a strong positive shift of the electron discrete energy level. This effect was not revealed by the previous calculations for critical solid clusters [11-13]. For positron in the $Ar_N$ clusters the self-compression leads to negative shift in energy. For positron in $Ar_N$ the $E_b^0$ term grows faster then the kinetic energy, therefore, $E_b^0$ bigger for self-compressed cluster then for rigid one. For $He_N$ this correlation breaks down. This is also reflected in the results for the critical positron clusters presented in Table 2.

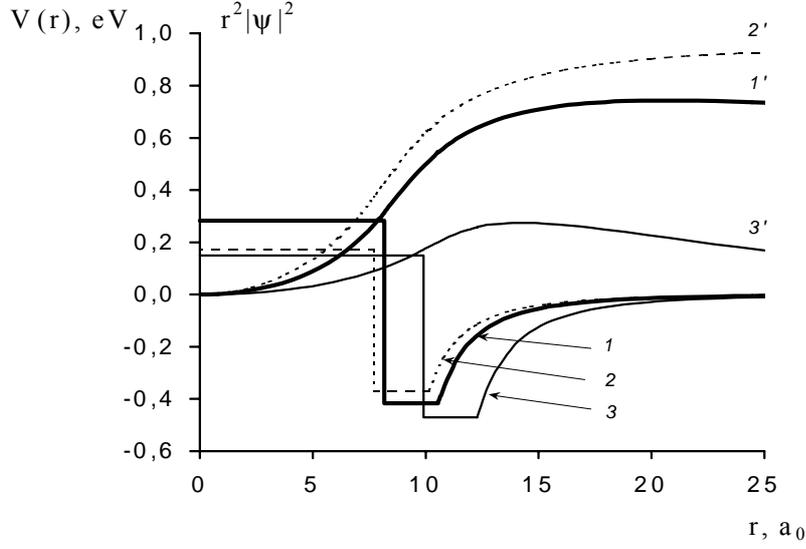

**Fig. 2**. The binding energy $E_b(N)$ for electron and positron in rigid and elastic clusters (dashed and solid line, respectively), calculated from Eq.(3): 1 - $Ar_N^+$; 2 - $Xe_N^-$; 3 - $Kr_N^-$; 4 - $He_N^+$. For $Xe_N^-$ the experiment gives $N^* =$ 5-8 [10].

## 4.2 Critical clusters

The critical size of cluster, corresponding to the number of atoms $N^*$, may be semi-quantitatively calculated from the Jost - Pais - Calogero (JPC) rule, i.e. from the condition for the appearance of the first bound state in the potential $V(r) < 0$, which is given by

$$\frac{m}{\hbar^2}\left|\int_0^\infty dr\, r V(r)\right| = I^*. \qquad (42)$$

For different potentials, usually employed in the nuclear physics, the value of $I^*$ changes from $\pi^2/8$ to 1.6 [42]. Solving equation (42) with respect to $R^*$, we can calculate $N^* = (R^*/\bar{r})^3$. It should be remembered that condition (42) was formulated for zeroth binding energy of captured particle.

As a rough estimation we have calculated $R^*$ and $N^*$ from (42) using square potential well. The results for rigid as well as for elastic clusters are presented in Table 2. According to this simple estimation the electronic stabilization must be observed for $N_{JPC}^* > 4$, 9 and 28 atoms for rigid and for $N_{JPC}^* > 6$, 13 and 78 atoms for elastic clusters of Xe, Kr and Ar respectively.

As we have mentioned above, the absence of the exchange (repulsive) interaction creates more favourable conditions for localization of the positron. Positron in critical $Ar_N$ clusters feels a deep potential well. Our estimations demonstrate the stabilization of positively charged clusters for $N_{JPC}^* > 4$, 18 and 19 for Ar, Ne and He, respectively. In these materials a small value of the derivative, $dV_0/d\rho$, causes that the self-compression does not influence significantly upon the binding energy of antiparticle. It is interesting to note the different influence of this effect on Ar as compared to Ne and He. Self-compressing leads to a positive shift of the positron energy level in $Ar_N$ and to a negative one for $Ne_N$ and $He_N$. This is determined by the competition between the size dependences of the bottom of the potential well, the polarization tail and the kinetic energy of antiparticle. In critical $Ne_N$ and $He_N$ clusters the polarizability tail is very small and positron encounters a nearly square potential well. Thus, our estimation of $N^*$ based on Eq.(42) and the square potential well is close to realistic values (Table 2).

In a second step, we determine the sizes of critical clusters $N^*$ by numerical solution of Eq.(37). To simplify the calculations we assume that $V_p(r)=0$ for $r>R+a$ [11]. Putting a=7R, which is good approximation because $|V_p(R+a)|<0.1\,\mathrm{meV}$, and owing to the fact that wave function in the region $r>R+a$ has a purely exponential form, we can replace the boundary condition outside the cluster from $r=\infty$ to $r=R+a$. The new boundary condition put at $r=R+a$ is

$$\frac{d}{dr}\ln u(r) = -\sqrt{\frac{2mE_b}{\hbar^2}}.$$

We determine the critical $N^*$ by calculating the least positive value of the binding energy. The results of calculations for $N^*$ and $E_b(N^*)$ are presented in Table 2. The actual forms of the pseudopotential (38) and the density $r^2|\psi(r)|^2$ for electron in Xe and Ar and positron in Ar critical clusters are plotted in Figs. 3.

It is interesting to compare the obtained values of $N^*$ with the ones calculated from (42). As one could surmise, for electronic clusters $N^* < N^*_{JPC}$. It is stipulated by the fact that condition (42) was derived for $V(r)<0$. However, in the electronic clusters $V(r)>0$ for r<R and $V(r)<0$ for r>R, i.e., in the interior of the cluster electron encounters not potential well but a barrier (see Figs. 3). Therefore, the attractive and repulsive part of V(r) compensate each other in Eq.(42). To fulfil the equality in Eq.(42) the "negative" region of V(r) has to be increased which is equivalent to a fictious increase of $N^*$. The results of numerical solution of Eq.(37) confirm the role of self-deformation. The magnitudes of $N^*$ for the rigid and elastic clusters differ on 30%. This is caused by significant magnitude of the derivative $dV_0/d\rho$. In spite of that in smallest clusters the size self-compression is neutralised by the size self-tension.

**Table 2.** *The number of atoms $N^*$ constituting the electron and positron critical clusters for different rare gases and two different binding potentials. The $N^*_{JPC}$ values were determine from rule (42), which corresponds $E_b=0$, using square potential barrier $(I^*=\pi^2/8)$. The given values of $E_b$, $N^*$ and $N_T$ are*

*determined quantum mechanically with potential given by (38). The values of $N_T$ correspond to $E_b = k_B T$.*

| | | $Xe_N^-$ | $Kr_N^-$ | $Ar_N^-$ | $Ar_N^+$ | $Ne_N^+$ | $He_N^+$ |
|---|---|---|---|---|---|---|---|
| Rigid | $N_{JPC}^*$ | 4-5 | 9-10 | 28-29 | 4 | 17 | 18-19 |
| | $N^*$ | 4 | 8 | 19 | 5 | 19 | 20 |
| | $E_b$, meV | 0.008 | 0.13 | 0.27 | 3.12 | 0.02 | 0.02 |
| | $N_T$ | 7 | 14 | 32 | 6 | 23 | 22 |
| Elastic | $N_{JPC}^*$ | 6-7 | 13-14 | 78-79 | 4 | 18 | 19 |
| | $N^*$ | 5 | 9 | 24 | 5 | 20 | 20 |
| | $E_b$, meV | 0.37 | 0.019 | 0.00002 | 4.33 | 0.07 | 0.0005 |
| | $N_T$ | 9 | 17 | 52 | 6 | 23 | 22 |

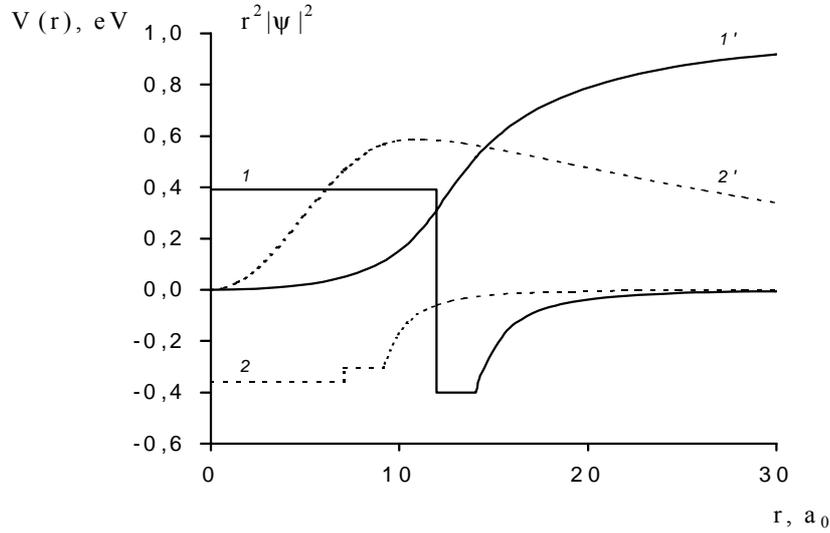

**Fig. 3**. Pseudopotential $V(r)$ (curves 1, 2, 3) and radial density distribution $r^2|\psi(r)|^2$ (arbitrary units, curves 1', 2', 3') for: a) Critical elastic $Xe_5^-$ (solid lines 1, 1'), critical rigid $Xe_4^-$ (dashed lines 2, 2') and "thermal" cluster $Xe_9^-$ (solid lines 3, 3'); b) Critical elastic $Ar_{24}^-$ (solid lines 1, 1') and critical elastic $Ar_5^+$ (dashed lines 2, 2').

For positron clusters the values of $N^*$ are higher then $N_{JPC}^*$. This result is clear too: despite of V(r)<0 for arbitrary r (see Fig. 3), the using $I^* = \pi^2/8$ in (42) corresponds to use of square potential well. For positron states the account for the size dependence of the polarization tail leads to $N^* > N_{JPC}^*$.

Thus, for the fixed magnitudes of the density $\rho_0$ and temperature T, the "optimal" clusters are: $Xe_{N>5}^-$, $Kr_{N>9}^-$, $Ar_{N>24}^-$, $Ar_{N>5}^+$, $Ne_{N>20}^+$, $He_{N>20}^+$. Note that these values of $N^*$ are underestimated, because they do not take into account the possibility of thermal excitations, i.e. $E_b > k_B T$. In a final step, we calculate the sizes of clusters $N_T$ corresponding to the condition $E_b = k_B T$ (see Table 2).

Analysing the results for elastic electronic clusters it is seen that the calculated critical number for $Xe_N^-$ agrees well with the experimental result giving $N^*$=5-8 [10]. On the other hand, the agreement with another theoretical result for solid Xe and Kr clusters ($N^*$=8 and 14) is quit good, but not for Ar ($N^*$=46) [13]. Our results point on the noticeable influence of self-compression, which has not been taken into account before. Self-deformation leads to increase of N by 30-50%. In view of the latter fact, the accurate prediction of critical number $N^*$ by the authors of Refs. 11-13 must be considered rather as fortuitous.

## 5. Conclusions

The estimation presented in this work demonstrate that analytical equation (3) points on smaller sizes of electron critical cluster and thus gives better agreement with measured values. The theory underlying this formula does not use adjustable parameters and is based on the information about electron/positron scattering length, and the Lennard-Jones potential. We have developed formal density-functional theory of a finite classical system which allows to account for the effect of self-compression, originating from the curvature of cluster surface, and the effect of self-tension due to reduction of the cluster's size. The critical sizes of clusters were determined quantum-mechanically by solving Schrodinger equation and from Jost-Pais-Calogero criterion. The effects of self-compression and tension give a significant contribution to the critical sizes of clusters charged by electron and should be taken into consideration in any comparison of critical cluster's sizes with the measured ones. For positron charged clusters the elastic effects are negligible.

Our model based on the continuum approximation may be not used to describe the localization of electron/positron at a single atom having a large polarizability. The appropriate methods for the solution this problem have been developed before [43,44] (see also Ref.45). However, direct application of these methods to the single "ion" $Xe_1+e^+$ needs knowledge of the radius of the short-range core of potential.

The behaviour of the one-particle potential V(r) of electronic clusters qualitatively resemble that for positron in a metal with negative positron work function (Al, Mo, Fe, Ni) [46]. It suggests a possibility of application of our method to large metallic clusters charged by positron. The results of this investigation may find application in positron diagnostics in ultradispersed media and possibly in rare-gas atom nanotechnology.

This work was supported, in part, by International Soros Education Program through the grant APU 072082.